\documentclass[aps,prl,twocolumn,twoside,floatfix,amsmath,showpacs,superscriptaddress]{revtex4-1}

\usepackage{graphicx}
\usepackage{natbib}
\usepackage{color}
\usepackage[normalem]{ulem}

\begin{document}

\newcommand{\CRS}{CeRu$_2$Si$_2$}
\newcommand{\YRS}{YbRh$_2$Si$_2$}
\newcommand{\YNP}{YbNi$_4$P$_2$}

\newcommand{\sovert}[1]{\ensuremath{#1\,\mu\textnormal{V K}^{-2}}}
\newcommand{\resist}[1]{\ensuremath{#1\,\mu\Omega\textnormal{cm}}}
\newcommand{\mJmolK}[1]{\ensuremath{#1\,\textnormal{mJ mol}^{-1} 
\textnormal{K}^{-2}}}
\newcommand{\lorenzunits}{\ensuremath{\,\textnormal{W \Omega K}^{-2}}}

\newcommand{\kB}{\ensuremath{k_{\textnormal{B}}}}
\newcommand{\gfac}{\ensuremath{g_{\textnormal{eff}}}}
\newcommand{\muB}{\ensuremath{\mu_{\textnormal{B}}}}

\newcommand{\ie}{{\em i.e.}}
\newcommand{\eg}{{\em e.g.}}
\newcommand{\etal}{{\em et al.}}

\newcommand{\replace}[2]{\sout{#1} \textcolor{red}{#2}}
\newcommand{\tred}[1]{\textcolor{red}{#1}}
\newcommand{\tblue}[1]{\textcolor{blue}{#1}}

\title{Cascade of magnetic field induced Lifshitz transitions in the ferromagnetic Kondo lattice material \YNP}

\author{H.~Pfau}
\affiliation{Max Planck Institute for Chemical Physics of Solids, D-01187 
Dresden, Germany}
\affiliation{Stanford Institute for Materials and Energy Sciences, SLAC National Accelerator Laboratory, 2575 Sand Hill Road, Menlo Park, California 94025, USA}
\author{R.~Daou}
\affiliation{Normandie Univ, ENSICAEN, UNICAEN, CNRS, CRISMAT, 14000 Caen, France.}
\author{S.~Friedemann}
\author{S.~Karbassi}
\affiliation{HH Wills Laboratory, University of Bristol, BS8 1TL Bristol, UK}
\author{S.~Ghannadzadeh}
\affiliation{High Field Magnet Laboratory, University of Nijmegen, 6525 ED Nijmegen, The Netherlands}
\author{R.~K\"uchler}
\author{S.~Hamann}
\author{A.~Steppke}
\author{D.~Sun}
\author{M.~K\"onig}
\affiliation{Max Planck Institute for Chemical Physics of Solids, D-01187 
Dresden, Germany}
\author{A.~P.~Mackenzie}
\affiliation{Max Planck Institute for Chemical Physics of Solids, D-01187 
Dresden, Germany}
\affiliation{Scottish Universities Physics Alliance (SUPA), School of Physics and Astronomy, University of St.\ Andrews, St.\ Andrews KY16 9SS, United Kingdom}
\author{K.~Kliemt}
\author{C.~Krellner}
\affiliation{Physikalisches Institut, Johann Wolfgang Goethe-Universit\"{a}t, 
D-60438 Frankfurt am Main, Germany}
\author{M.~Brando}
\affiliation{Max Planck Institute for Chemical Physics of Solids, D-01187 
Dresden, Germany}

\date{\today}


\begin{abstract}

A ferromagnetic quantum critical point is thought not to exist in two and three-dimensional metallic systems yet is realized in the Kondo lattice compound YbNi$_4$(P,As)$_2$, possibly due to its one-dimensionality. It is crucial to investigate the  dimensionality of the Fermi surface of YbNi$_4$P$_2$ experimentally but common probes such as ARPES and quantum oscillation measurements are lacking. Here, we studied the magnetic field dependence of transport and thermodynamic properties of \YNP. The Kondo effect is continuously suppressed and additionally we identify nine Lifshitz transitions between 0.4 and 18\,T. We analyze the transport coefficients in detail and identify the type of Lifshitz transitions as neck or void type to gain information on the Fermi surface of YbNi$_4$P$_2$. The large number of Lifshitz transitions observed within this small energy window is unprecedented and results from the particular flat renormalized band structure with strong $4f$-electron character shaped by the Kondo lattice effect.
\end{abstract}

\maketitle

The Fermi surface (FS) topology plays a key role in understanding metallic materials, because their electronic properties are determined by thermally excited quasiparticles confined to a narrow window around the Fermi energy. Angle-resolved photoemission spectroscopy (ARPES) and quantum oscillation (QO) measurements are the most common tools to determine the FS. While ARPES relies on an excellent surface quality, QOs need to be performed at high magnetic fields of the order of 10\,T in metals but are typically interpreted using band structure calculations at zero field. The ability of QOs to interpret zero field properties is therefore under intense discussion, \eg\ in high-temperature superconductors~\cite{liu_2010,leboeuf_2011,doiron-leyraud_2007,barisic_2013,grissonnanche_arxiv} and low-carrier-density topological materials with surface states~\cite{xu_2015}.

These considerations are especially relevant to Kondo lattice systems in which local $f$-electrons and conduction electrons form composite heavy quasiparticles below the Kondo temperature $T_{\textrm{K}}$. These systems develop flat bands close to the Fermi level and van-Hove singularities in the renormalized density of state (DOS) due to the coherence effects in the lattice \cite{hewson_kondo}. The Kondo energy scale \kB$T_{\textrm{K}}$ is a measure of the Fermi energy of heavy fermion systems. Since it roughly corresponds to a Zeeman energy $\frac{1}{2}g_{\textrm{eff}}$\muB$B$ for magnetic fields around 10\,T, they are very susceptible to FS changes due to magnetic field induced Lifshitz transitions (LTs): changes in the topology of the FS without symmetry breaking. \cite{lifshitz_1960}. It is particularly difficult to predict the exact field strengths at which those LTs will take place because of strong correlations and the specific crystalline electric field (CEF) ground state \cite{zwicknagl_2011}.

LTs are an integral part of the complex  phase  diagram  of  correlated  materials and
have been reported in heavy fermion (HF) compounds such as \YRS\ \cite{tokiwa_2005,rourke_2008,pfau_2013_prl,pourret_2013} and CeIrIn$_5$ \cite{aoki_2016}, near the metamagnetic transition in \CRS\ \cite{daou_2006a,pfau_2012,boukahil_2014} and in the hidden ordered phase of URu$_2$Si$_2$ \cite{altarawneh_2011,malone_2011}. They are also discussed in connection with superconductivity, \eg\ in certain ferromagnets \cite{kotegawa_2011,yamaji_2007,bastien_2016}, in URhGe \cite{yelland_2011,gourgout_2016}, in Sr$_2$RuO$_4$ \cite{steppke_2017}, in high-temperature superconductors \cite{liu_2010,leboeuf_2011} and in topological systems, for example in Dirac semimetals~\cite{xu_2015}. 

In this letter we study \YNP, which has a quasi-1D crystal structure with isolated chains of magnetic Yb$^{3+}$ atoms along the crystallographic $c$-axis \cite{deputier_1997}. The reported resistivity anisotropy hints towards a 1D character of the electronic structure \cite{steppke_2013}. Uncorrelated band structure calculations with dominating Ni-3$d$  DOS predict two flat FS sheets \cite{krellner_2011}. \YNP\ is a Kondo lattice with $T_{\textrm{K}}=8$\,K, which orders ferromagnetically (FM) at $T_{\textrm{C}} \approx 0.15$\,K with a small ordered moment of 0.05$\mu_\mathrm{B}$ aligned within the $(a,b)$ plane. While the FM state is suppressed at $B_{\textrm{c}} \approx 0.06$\,T applied along the $c$-axis \cite{steppke_2013}, \YNP\ can be tuned towards a ferromagnetic quantum critical point (QCP) by As-substitution \cite{krellner_2011,steppke_2013}. Such a FM QCP was thought not to exist in metallic systems for dimensions $d\geq 2$ \cite{brando_2016,belitz_1999} and is believed to be realized in \YNP\ due to its 1D-character \cite{steppke_2013}. It is therefore crucial to experimentally determine the FS and verify its low-dimensional character. 

Difficulties with cleaving \YNP~crystals hampered ARPES measurements; QOs  are unavailable to date and additionally suffer from problems measuring the zero-field FS. Recent studies on \YRS\ combine magnetic field dependent thermopower, resistivity and magnetostriction measurements to form a powerful tool set that detects changes in the FS due to field induced LTs~\cite{pfau_2013_prl,pourret_2013}. It also allows studying magnetic field ranges below the fields necessary for QO measurements. The type of FS changes in YbRh$_2$Si$_2$ was successfully compared with renormalized band structure calculations \cite{zwicknagl_1992,zwicknagl_2011,pfau_2013_prl} but such calculations are unavailable for \YNP. Therefore, we extended our analysis of the thermopower and resistivity to compare observed signatures to general theoretical predictions for transport coefficients close to a LT. Our analysis enables us to identify not only the magnetic field of the LTs, but also to determine their topological character and carrier type. Hence, our method provides detailed information about the FS of \YNP, where standard methodology fails.

We show how a relatively small external magnetic field dramatically modifies the FS of the HF system \YNP\ producing in total nine LTs that we analyze in detail. Similar to \YRS, the topological changes are superimposed on a continuous suppression of the Kondo effect with increasing field \cite{pfau_2013_prl,naren_2013}. Our study indicates that the observation of several Zeeman-driven LTs on top of a smooth suppression of the Kondo effect in magnetic field is a generic property of Kondo lattice systems. Additionally, the behavior of \YNP~in finite magnetic field hints towards a spin density wave scenario for the QCP tuned by chemical pressure in YbNi$_4$(P,As)$_2$.

\begin{figure}[t]
  \begin{center}
    \includegraphics[width=\linewidth]{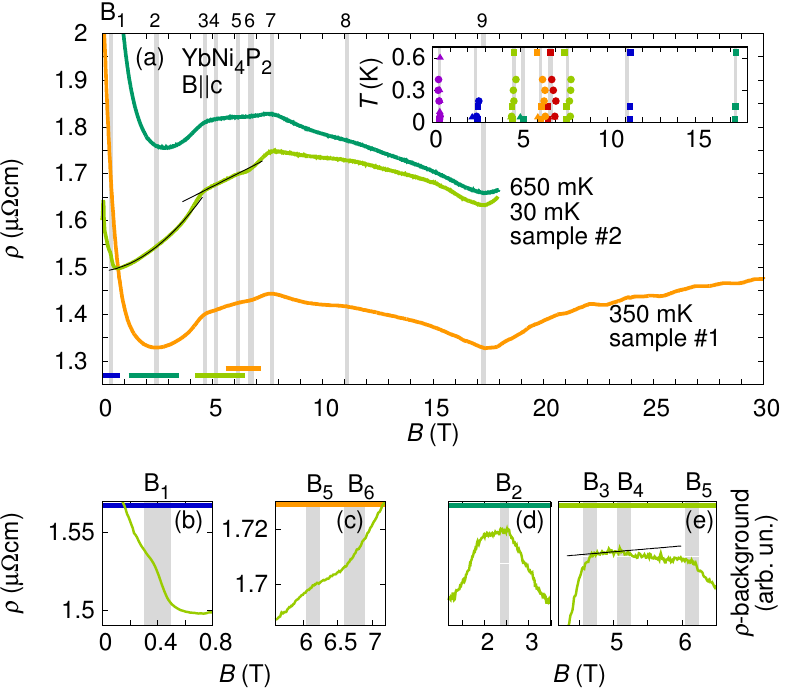}
  \end{center}
	  \caption{Resistivity. (a) Resistivity $\rho(B)$ as a function of magnetic field.(b), (c) A zoom into the region around $B_1$ and $B_{5,6}$, respectively. (d), (e) Background subtracted resistivity to highlight the signatures at $B_{2,4}$. The background determined from a linear (e) or quadratic (d) fit to the data is shown in (a) as dashed lines. Dashed line in (e) is a guide to the eye to highlight the changes around $B_4$. The field intervals for (b)-(e) are marked with solid bars on the bottom of (a). Inset: $T$-dependence of transition fields from $\rho$ of sample \#2 (squares) and \#3 (circles), and $\lambda$ of sample \#4 (triangles). Gray vertical lines represent the transition fields $B_i$, $i = 1\dots 9$, their thickness corresponds to the error of $B_i$.}

  \label{fig:rho}
\end{figure}

Our measurements on single crystalline samples \cite{kliemt_2016} focus on a magnetic field $B\parallel c$ above $B_c$, which suppresses the ferromagnetic order. We performed resistivity measurements on two samples with current $I \parallel c$. Sample \#1 has a residual resistivity $\rho_{0} \approx 1$\resist\, and was measured in magnetic fields up to 30\,T at the High Field Magnet Laboratory in Nijmegen. Sample \#2 with $\rho_{0} \approx $1.7\resist\, was cut from sample \#4 and shaped into a thin wire using a focussed ion beam patterning. It was used for resistivity measurements in a dilution refrigerator down to 30\,mK and in fields up to 18\,T. We checked that the FIB patterning did not alter the resistivity of the sample. Sample \#3 with $\rho_{0} \approx 2.6$\resist\, was used for thermal transport measurements up to 12\,T using a standard one-heater-two-thermometer configuration. Magnetostriction was measured up to 10\,T on the largest sample \#4 (length $L$ = 2\,mm) by means of a high-resolution capacitive CuBe dilatometer \cite{kuechler_2012}.

\begin{figure}[t]
	\begin{center}
		\includegraphics[width=\linewidth]{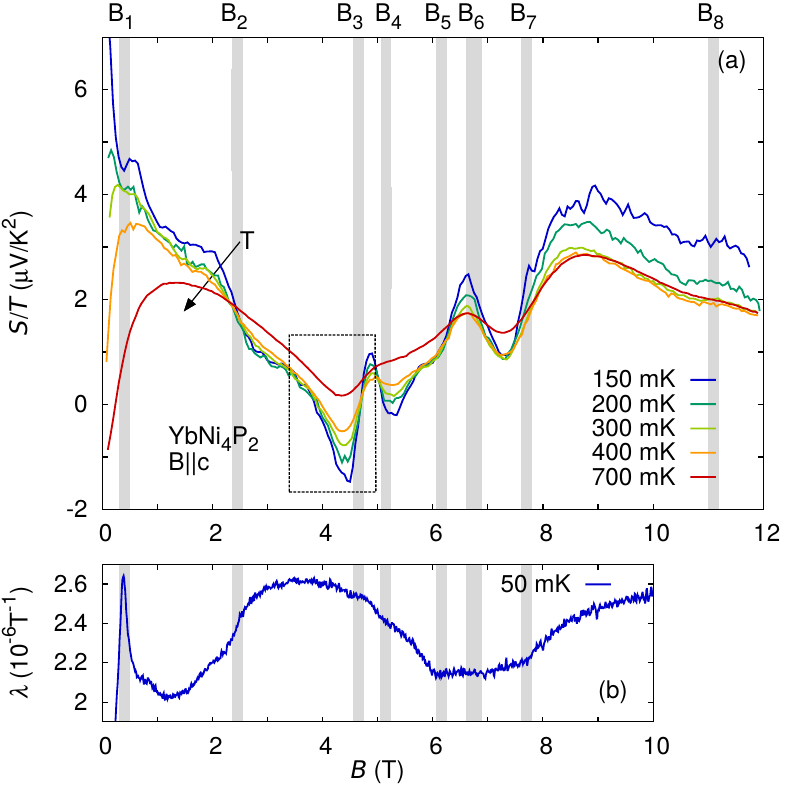} \\	
	\end{center}
	\caption{Thermopower and magnetostriction. (a) Thermopower of sample \#3 plotted as $S(B)/T$ as a function of magnetic field. The box highlights the signature around $B_3$ which we analyze in detail. (b) Linear magnetostriction coefficient $\lambda(B)$ of sample \#4 as a function of $B$. Gray vertical lines represent the transition fields $B_i$, $i=1\dots 9$, their thickness corresponds to the error of $B_i$.}
	\label{fig:S}
\end{figure}

The resistivity $\rho(B)$ is shown in Fig.~\ref{fig:rho}(a), Fig.~\ref{fig:S}(a) presents the thermopower as $S(B)/T$ and Fig.~\ref{fig:S}(b) the magnetostriction coefficient $\lambda(B)=\partial(\Delta L(B)/L)/\partial B$ for length changes along the $c$-axis.

At small magnetic fields $B \leq 1$\,T, we observe a negative magnetoresistance, which is typical for Kondo systems. It indicates the suppression of spin-flip scattering and hence a suppression of the Kondo effect. The thermopower $S(B)/T$ varies strongly with temperature in this field range, which can be related to the strong fluctuations in the vicinity of the QCP in YbNi$_4$(P,As)$_2$. Moreover, $\lambda(B)$ changes sign across $B_{\textrm{c}} \approx 0.06$\,T \cite{steppke_2013}, which is a clear signature of a symmetry breaking phase transition in a Yb-based Kondo-lattice system \cite{garst_2005}.

We focus on the signatures in all three quantities above $B_{\textrm{c}} \approx 0.06$\,T. Since all quantities show a rich magnetic-field dependence, we use the following strategy to identify in total nine transition fields that we list in Tab.~\ref{tab:lifshitz}. 1) We assign a transition to magnetic fields, where we can unambiguously observe a kink in either $\rho(B)$ (see also Fig.~\ref{fig:rho}(b,c)) and/or $\lambda(B)$. These fields are $B_1$, $B_6$ and $B_7$ ($\lambda$, $\rho$); $B_{3}$, $B_8$ and $B_9$ ($\rho$); and $B_{5}$ ($\lambda$). 2) We assign a transition to every field, were we observe weak signatures, but in all three quantities -- a kink in $\lambda$ and $\rho$ (see also Fig.~\ref{fig:rho}(d,e)) and a $T$-independent crossing in $S/T$. These field are $B_{2}$ and $B_4$.

Importantly, the position of the transition fields is temperature independent (see inset Fig.~\ref{fig:rho}(a)). Additionally, $\lambda$ always stays positive for $0.06 < B \leq 10$\,T and does not change sign, which rules out further symmetry-breaking transitions. Both observations suggest the presence of LTs at these fields \cite{lifshitz_1960,blanter_1994}. This finding is corroborated by results of specific heat measurements: Except for the ferromagnetic phase transition at 150\,mK ($B=0)$, there is no sign for another finite temperature phase transition at higher fields ($B\parallel c$) in the temperature dependence of the specific heat, measured between 60\,mK and 4\,K for several fields up to 12\,T \cite{sun_unpublished}. Furthermore, no significant enhancement of the magnetization $M(B)$ is observed across $B_i$ (not shown).
\begin{table}[bh]
 \centering
\begin{tabular}{@{}*{10}{c}@{}}
\toprule
& $B_1$ & $B_2$ & $B_3$ & $B_4$ & $B_5$ & $B_6$ & $B_7$ & $B_8$ & $B_9$\\
\hline
$B (T)$ & 0.40 & 2.45 & 4.65 & 5.15 & 6.15 & 6.7 & 7.70 & 11.0 & 17.5\\
\toprule 
\end{tabular}
\caption{Magnetic field values of the LTs. The error of $B_i$ is 0.1\,T.}
\label{tab:lifshitz}
\end{table}

\begin{figure}[t]
	\begin{center}
		\includegraphics[width=1.\linewidth]{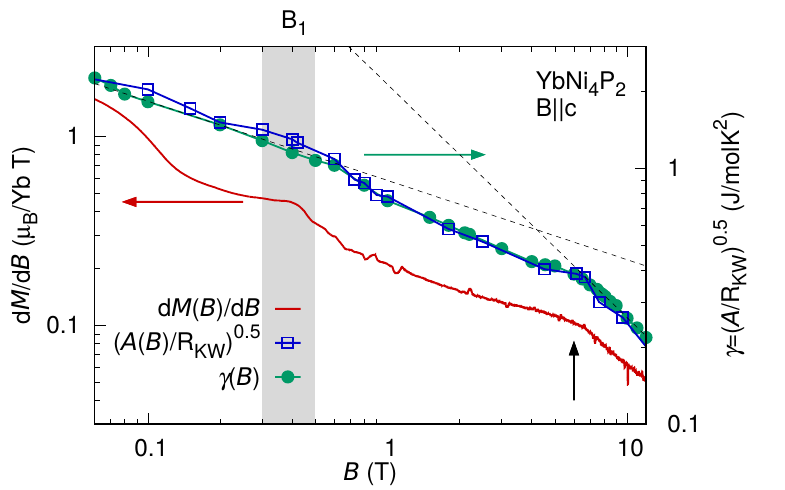}
	\end{center}
	\caption{Field dependence of the effective mass above $B_c$. As a measure of the effective mass, we plot here the field derivative of the magnetization $\mathrm{d}M/\mathrm{d}B$, the specific heat coefficient $\gamma(B)$ and the square root of the $A$ coefficient of the resistivity in a double-logarithmic plot. The dashed lines highlight the change of slope in $\gamma(B)$ just above $B_1$ (gray bar) and around $B_5<B<B_7$ (short arrow). The changes around the latter field scale are too broad in our measurements to be connected to a single $B_i$.}
	\label{fig:m*}
\end{figure}

To investigate, if the ground state of \YNP\ is a Fermi liquid, we measured the temperature dependence of $\rho(T)$, which indeed follows $\rho(T) = \rho_{0} + AT^{2}$ at all $B_{i}$ (not shown). This indicates that the LTs are not associated with anomalous or quantum critical behavior, as sometimes observed in metamagnetic systems like CeRu$_{2}$Si$_{2}$ \cite{daou_2006a}. 
Having established the Fermi liquid ground state, we can study the field evolution of the effective mass $m^*$ using the relation $m^{*}(B) \propto \sqrt{A}(B) \propto \chi(B) \propto \gamma(B) \propto \mathrm{DOS}(B)$. Here, $A$ is the quasiparticle--quasiparticle scattering rate extracted from the $T^2$ term in $\rho(B)$, $\chi = \mathrm{d}M/\mathrm{d}B$ is the magnetic susceptibility extracted from magnetization measurements,  $\gamma(B)$ the Sommerfeld coefficient. All four quantities should be proportional to the DOS. Fig.~\ref{fig:m*} shows $\gamma(B)$, $\chi(B)$ and $\sqrt{A(B)/\mathrm{R_{KW}}}$, where $\mathrm{R_{KW}}$ is the Kadowaki-Woods ratio, which we determined to be 2\,$\mathrm{\mu\Omega\,cm/(J/molK^2)^2}$ in \YNP. All three quantities demonstrate, that $m^{*}$ decreases strongly but continuously between 0.06\,T and 10\,T. Above 10\,T, $\gamma$ is still about 0.2\,J/molK$^{2}$, which confirms the persistence of the Kondo lattice effect even at this high field. Similar behavior was observed in \YRS\ \cite{pfau_2013_prl}. Interestingly, $m^{*}$ shows significant changes of slope only at certain $B_{i}$, i.e., at $B_{1}$ and around $B_{5} < B < B_{7}$.

\begin{figure}[t]
	\begin{center}
		\includegraphics[width=1.\linewidth]{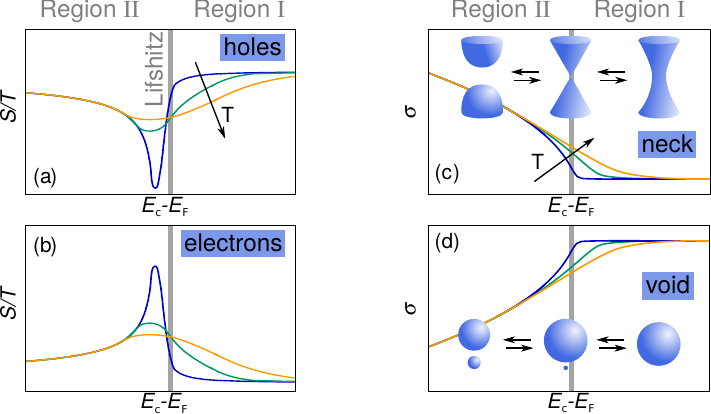}
	\end{center}
	\caption{Lifshitz transitions. Panels (a)-(d) present theoretical calculations of the signatures in thermopower $S$ and conductivity $\sigma$ close to a Lifshitz transition. The plots are reproduced from \cite{blanter_1994,varlamov_1989} for the clean case and presented for three different temperatures. The sign of the thermopower signature (maximum or minimum) is defined by the type of charge carrier. The sign of the conductivity, (i.e., minimum or maximum in $\partial^2 \sigma / \partial (E_\mathrm{c}-E_\mathrm{F})^2$), is defined by the type of Lifshitz transition, i.e., either neck or void type (see main text)\cite{varlamov_1989}.}
	\label{fig:theory}
\end{figure}

In the following we want to compare our experimental results with theoretical predictions for $\rho$ and $S$ close to a LT. There are two main types of LTs as displayed in Fig.\ref{fig:theory}(c),(d): the void type where a FS sheet vanishes, and a neck type where a FS splits into two sheets. Following the terminology of Ref.~\cite{blanter_1994,varlamov_1989}, the side of the transition where the new pocket is absent and where the neck is not broken corresponds to region I. Figures \ref{fig:theory}(a)-(d) present theoretical predictions for the signatures one expects to observe in electrical conductivity $\sigma$ and thermopower at a LT of a three-dimensional band \cite{blanter_1994,varlamov_1989,buhmann_2013}. $E_\mathrm{c}-E_\mathrm{F}$ defines the distance of the extremum in the bandstructure to the Fermi energy. Considering Zeeman-driven Lifshitz transitions, this can be translated into the experimental parameter magnetic field using $E=g\mu_\mathrm{B}B$. The signatures in $\sigma$ and $S/T$ become smeared with increasing temperature; however the position of the transition is $T$-independent.
Such signatures were observed experimentally across LTs in several different systems, e.g., in elements and metallic solid solutions \cite{bruno_1994,varlamov_1989}, semiconductors \cite{meng_2003,varlamov_1989} and high-temperature superconductors \cite{hodovanets_2014,shen_2016}. However, these materials need to be tuned to the LT by external pressure or doping. In most cases this corresponds to a much higher energy shift compared to the shift due to magnetic fields of the order of a few tesla, which is sufficient for the flat bands of the Kondo lattice to undergo a LT.

In contrast to thermodynamic quantities, transport properties such as resistivity and thermopower are most affected by changes in the scattering time and not the DOS close to a LT and usually show a stronger response \cite{blanter_1994}. This response is asymmetric around $E_\mathrm{c}-E_\mathrm{F} =0$ since a new scattering channel appears on one side of the LT and it is absent on the other. The extremum in the thermopower is located slightly away from the LT: For $T\neq0$ scattering into the not yet born FS is already possible \cite{varlamov_1989}.
Using these transport properties, one can in principle determine the type of LT (using the conductivity $\sigma$), the carrier type involved (using $S/T$), and its direction, i.e., which side of the transition corresponds to region I and which one to region II  (from asymmetry in $S/T$) \cite{varlamov_1989}. In 3D, these signatures are independent of the specific band structure \cite{varlamov_1989}, but differ for lower dimensions \cite{blanter_1994}.

The transition $B_3$ follows these theoretical predictions. We can extract detailed information about the type of LT comparing Fig.\ \ref{fig:theory}(c),(d) with Fig.\ \ref{fig:rho}(a) and Fig.\ \ref{fig:theory}(a),(b) with Fig.\ \ref{fig:S}(a) (see highlighted region around $B_3$).
We assume $\sigma = 1/\rho$ and a negligible contribution from the Hall resistivity for simplicity. The results of this comparison for $B_3$ are: i) The peak in $S/T$ is at fields smaller than $B_{3}$, corresponding to region II; ii) the peak is negative, suggesting that hole carriers are dominant; iii) the slope of $\rho(B)$ decreases for $B > B_{3}$, so the transition is of neck type. Hence a neck joins two pockets of a hole band as the field increases, crossing $B_3$. All other transitions also show kinks in the resistivity which can be analyzed in the same scheme. From this analysis we propose that all transitions are of neck type, beside $B_6$ and $B_9$ which are presumably of void type. The corresponding thermopower signatures show similarities to the theoretical prediction. However, they are hard to interpret for one or more of the following reasons. They are 1) covered by a strongly $B$-dependent background, 2) lie very close to each other, 3) cannot be accessed with the limited field range of the thermopower measurement, or 4) show different signatures in $S/T$ than theoretically predicted.

Three of the LTs ($B_1$, $B_5$, $B_7$) show a strong response in thermodynamic quantities (see Fig.\ \ref{fig:S}(b) and \ref{fig:m*}). This signals a Zeeman-splitting of a strong DOS feature that moves through $E_\mathrm{F}$. In \YRS, the spin-spitting of the Kondo resonance causes such an effect. Its field scale corresponds to the Kondo energy scale. In \YNP, $B_5$ and $B_7$ may have the same origin. However, large thermodynamic effects can also indicate that low-dimensional FSs are involved. The DOS gradually drops as $E^{1/2}$ towards the band edge for a parabolic band in 3D, it is energy-independent in 2D and diverges as $E^{-1/2}$ in 1D. Concrete calculations of the specific heat close to a LT in 2D predict stronger signatures compared to the 3D case \cite{blanter_1994}. We expect quantities related to the DOS to show the strongest signatures for a 1D LT.

The uncorrelated band structure calculations of \YNP\ predict two quasi-1D FS sheets in the $k_x$-$k_y$ plane due to the one-dimensional character of the crystal structure \cite{krellner_2011}. These flat sheets can also undergo a LT. Hence, $B_{5}$ and $B_{7}$, but especially $B_1$ may be connected to LTs in the renormalized quasi-1D FS sheets of \YNP.

Our results have also implications for the controversial topic whether the Fermi volume loses the $f$-electron right at the QCP (Kondo breakdown scenario)  or well within the  magnetically  ordered  state (spin density wave scenario)~\cite{si_2001,gegenwart_2008,woelfle_2011,kummer_2015,paschen_2016}. \YNP\ is located slightly on the magnetically ordered side of the pressure-induced QCP~\cite{steppke_2013}, in a regime where the Kondo breakdown and the spin density wave scenario make opposite predictions. Our results demonstrate that the FS is extremely sensitive to small external fields of the order of 1\,T. This implies the presence of weakly dispersing bands, which are shifted by the Zeeman splitting on a significant portion of the Brillouin zone. This is a strong indication that the $f$-degrees of freedom are involved in the formation of the FSs in the field range $B>B_c$, which was investigated in this study. Thus our results hint towards the spin density wave scenario for the pressure induced QCP. 

In conclusion, we have investigated the Kondo lattice system \YNP\ in magnetic fields above its ferromagnetic order. We discovered in total nine field induced LTs between 0.4\,T and 18\,T. We present an analysis method of transport properties, which allows us to identify the specific type of LTs being of void or neck type. This method enables us also to identify a hole band in which two pockets join in a neck transition across one of the LTs. We also find indications for the existence of FSs with a lower dimension, which is an important step towards an understanding of the ferromagnetic QCP in \YNP. Our analysis yields information about the bandstructure and its changes without involving specific band structure calculations and hence serves as a benchmark for future theoretical models such as renormalized band structure calculations.

The large number of Lifshitz transitions in a small magnetic field range reveals the presence of many extrema in the band structure of \YNP\ very close to the Fermi level, shaped by the Kondo lattice effect with anisotropic momentum-dependent hybridization acting in a multiband system. The magnetic field scale of the transitions is therefore to first order determined by the Kondo temperature $T_\mathrm{K}=8$\,K and to second order by the specifics of the hybridization and the multiband character. The comparison to other Kondo lattice systems such as \YRS\ suggests that this is a generic property of heavy fermion systems.

\begin{acknowledgments}
 We are indebted to C. Geibel, A. Gourgout, B. Schmidt, F. Steglich and A. A. Varlamov for useful discussions. This work was supported by the German Research Foundation (DFG) through the grants BR4110/1-1, KR3831/4-1, KU3287/1-1, and Fermi-NESt. H.P. acknowledges the support from the Alexander von Humboldt Foundation. S.F. and S.K. acknowledge support from the EPSRC under grant EP/N01085X/1.
\end{acknowledgments}

\bibliography{pfau_2016_prl}

\begin{thebibliography}{48}
\bibitem [{\citenamefont {Sun}\ and\ \citenamefont
  {et~al}()}]{sun_unpublished}%
  \BibitemOpen
  \bibfield  {author} {\bibinfo {author} {\bibfnamefont {D.}~\bibnamefont
  {Sun}}\ and\ \bibinfo {author} {\bibnamefont {et~al}},\ }\href@noop {}
  {}\bibinfo {note} {To be published}\BibitemShut {NoStop}%
\end{thebibliography}

\end{document}